\documentclass{article}
\usepackage{amsxtra,amssymb,amsmath,bm, color}
\def\Ref#1{(\ref{#1})}

\def\ri{\mathrm{i}}
\def\rd{\mathrm{d}}
\def\rD{\mathrm{D}}

\begin{document}
\noindent{\large\textbf{Decoherence in quantum systems in a static gravitational field}}

\vspace{2\baselineskip}\noindent
{\sffamily Ahmad~Shariati$^*$\footnote{e-mail: shariati@mailaps.org},
Mohammad~Khorrami$^*$\footnote{e-mail: mamwad@mailaps.org},
Farhang~Loran$^\dagger$\footnote{e-mail: loran@cc.iut.ac.ir}}

\vspace{\baselineskip}\noindent
$^*$ {\small Department of Physics, Alzahra University, Tehran 1993891167, Iran}\\
$^\dagger$ {\small Department of Physics, Isfahan University of Technology,
Isfahan 84156-83111, Iran}

\vspace{2\baselineskip}
\noindent{\textbf{Abstract}}\\
A small quantum system is studied which is a superposition of states localized in
different positions in a static gravitational field. The time evolution of
the correlation between different positions is investigated, and it is seen
that there are two time scales for such an evolution (decoherence).
Both time scales are inversely proportional to the red shift difference
between the two points. These time scales correspond to decoherences
which are linear and quadratic, respectively, in time.
\section{Introduction}
Interaction of a system with its environments could result in the so called
quantum decoherence, a decrease in the correlation between different parts
of the system. Such interactions, could also result in increasing
the correlation.

An example is the effect of a gravitational field on such correlations.
Gravitational time dilations, which depend on the position, cause the phase change
in different parts of the system located at different positions to be different.
That could result in a decoherence, or in certain cases a coherence, that is
an increase in the correlation between different parts.

Recently, this phenomena has been studied in \cite{NP} explicitly for a system of
$N$ harmonic oscillators in thermal equilibrium. An interested reader may
consult \cite{NP} and \cite{NP1} for a comprehensive list of references,
and additional arguments in favor of this observation.

Of course time dilation is not the only source of decoherence. Interaction
with environment is another source. These \textit{decoherence due to scattering},
occur at rates which are proportional to the square of the distance.
Some typical rates have been given in \cite{JZ} and \cite{Tg}, for example.

A quantum system in a gravitational field could affect the coherence properties
of the gravitational field as well. An example is discussed in chapter 4 of \cite{DA}.

Here a general quantum system is investigated, which is in a static gravitational field,
and is delocalized. The system is assumed to be a small system (of a small size
$\ell$) with internal degrees of freedom, say a molecule. By delocalization,
it is meant that the state of the system is a superposition of states at different
points of the space, distributed in a region of size $\Delta x$. The length scale
corresponding to the gravitational field (the curvature length) is $L$.
It is assumed that
\begin{align}
\ell&\ll L,\,\Delta x.
\intertext{$\Delta x$ could be smaller or larger than $L$, but more explicit calculations are
performed for the case}
\Delta x&\ll L.
\end{align}
It is seen that if the system is delocalized, the correlation between different parts of
it located in different positions varies with time. Two time scales are obtained for
these changes. Both are inversely proportional to the gravitational potential
difference (or the red shift difference) between the two points. The first time scales
corresponds to a change of the correlation which is linear in time, and could
be either an increase or a decrease. This time scale is finite only if there are
non real correlations between different points. The second time scale corresponds to
a decrease in the correlation, which is quadratic in time. This time scale is inversely
proportional to the energy fluctuations of the system. This is similar to the one obtained
in \cite{NP}.

The scheme of the paper is the following. A system is studied which has
internal degrees of freedom and is delocalized in a static gravitational field.
In section 2 the Hamiltonian, and in section 3 the time evolution of such a system
are investigated. In section 4 the results are specialized to the case the
system is in thermal equilibrium. Section 5 is devoted to the concluding remarks.
\section{The Hamiltonian corresponding to internal degrees of freedom}
The line element of a static gravitational field can be written as
\begin{equation}
\rd s^2=-c^2\,f^2\,\rd t^2+g_{i\,j}\,\rd x^i\,\rd x^j,
\end{equation}
where $f$ and $g_{i\,j}$'s depend on only $x^i$'s (the spatial coordinates).
An observer at rest at the point $x$ would measure the proper time $\tau$ which is related to $t$ as
\begin{equation}
\rd\tau=[f(x)]\,\rd t.
\end{equation}
A localized system at rest at the point $x$ could have energy levels $E_n$,
according to the observer at the same point $x$. The Hamiltonian corresponding
to these energy levels is the generator of the evolution of the proper time according
to that observer. Let us denote that Hamiltonian with $H$. For a system
at rest at the point $x$, the evolution operator would be
\begin{align}
U&=\exp\left(\frac{\tau\,H}{\ri\,\hbar}\right),\nonumber\\
&=\exp\left\{\frac{t\,[f(x)]\,H}{\ri\,\hbar}\right\}.
\end{align}
This suggests a global Hamiltonian $\bm{H}$, which is the generator of the evolution
of the time coordinate:
\begin{equation}\label{gd.4}
\bm{H}=[f(X)]\,H,
\end{equation}
where $X$ is the position operator:
\begin{equation}
\bm{H}\,|x,n\rangle=[f(x)]\,E_n\,|x,n\rangle,
\end{equation}
where $n$ labels the internal degrees of freedom.

A slightly more general version of \Ref{gd.4} is
\begin{equation}
\tilde{\bm{H}}=[f(X)]\,H+V(X).
\end{equation}
That change could be interpreted as changing the reference of energy of
the internal degrees of freedom by a position-dependent amount, or equivalently
adding a potential energy (corresponding to a force). It will be seen, however,
that this change does not affect the decoherence.
\section{The evolution}
Consider a density matrix
\begin{equation}
\rho(t)=\sum_{n,n'}\int\rd x\,\rd x'\,\rho(x,x';n,n';t)\,|x,n\rangle\langle x', n'|.
\end{equation}
Evolving this density matrix with a global Hamiltonian results in
\begin{align}
\rho(x,x',n,n';t)&=\rho(x,x',n,n';0)\nonumber\\
&\quad\times\exp\left\{
\frac{[E_n\,f(x)+V(x)-E_{n'}\,f(x')-V(x')]\,t}{\ri\,\hbar}\right\}.
\end{align}
Tracing this over the internal degrees of freedom,
\begin{align}
\tilde A(x,x';t)&=\sum_n\rho(x,x',n,n;t),\\
\intertext{it is seen that}
\tilde A(x,x';t)&=\exp\left\{\frac{[V(x)-V(x')]\,t}{\ri\,\hbar}\right\}\,A(x,x';t),\\
\intertext{where}
A(x,x';t)&=\sum_n\rho(x,x',n,n;0)\,\exp\left\{\frac{[f(x)-f(x')]\,E_n\,t}{\ri\,\hbar}\right\}.
\end{align}
A measure of the decoherence of the system is a decrease of $|\tilde A|$ in time.
It is seen that
\begin{equation}
|\tilde A(x,x';t)|=|A(x,x';t)|.
\end{equation}
Hence $V(X)$ does not affect the decoherence. Defining $\gamma$ and $a_n$'s and $\chi_n$'s through
\begin{align}
\gamma(x,x';t)&=\frac{A(x,x';t)}{A(x,x';0)},\\
\rho(x,x',n,n;t)&=\tilde A(x,x';t)\,a_n(x,x';t)\,\exp[\ri\,\chi_n(x,x';t)],
\intertext{where $a_n$'s are real and nonnegative and $\chi_n$'s are real, it is seen that}
1&=\sum_n\,a_n(x,x';t)\,\exp[\ri\,\chi_n(x,x';t)],\\
\gamma(x,x';t)&=\sum_n a_n(x,x';0)\,\exp[\ri\,\phi_n(x,x';t)],\\
\intertext{where}\label{gd.19}
\phi_n(x,x';t)&=\chi_n(x,x';0)-\frac{[f(x)-f(x')]\,E_n\,t}{\hbar}.
\end{align}
If $E_n$'s are commensurate, then $\gamma(x,x';t)$ would be periodic in time,
with the period $t_0$ satisfying
\begin{equation}
t_0=\frac{h}{[f(x)-f(x')]\,\varepsilon},
\end{equation}
where $\varepsilon$ is the greatest common divisor of $E_n$'s.

Up to second order in time,
\begin{align}
\gamma(x,x';t)&=1+\frac{[f(x)-f(x')]\,t}{\ri\,\hbar}\,
\sum_n E_n\,a_n(x,x';0)\,\exp[\ri\,\chi_n(x,x';0)]\nonumber\\
&\quad+\frac{1}{2}\,\left\{\frac{[f(x)-f(x')]\,t}{\ri\,\hbar}\,\right\}^2\,
\sum_n (E_n)^2\,a_n(x,x';0)\,\exp[\ri\,\chi_n(x,x';0)]\nonumber\\
&\quad+\cdots
\end{align}
So,
\begin{align}
|\gamma(x,x';t)|^2&=1+\frac{2\,[f(x)-f(x')]\,t}{\hbar}\,
\sum_n E_n\,a_n(x,x';0)\,\sin[\chi_n(x,x';0)]\nonumber\\
&\quad+\left\{\frac{[f(x)-f(x')]\,t}{\hbar}\right\}^2\,
\left|\sum_n E_n\,a_n(x,x';0)\,\exp[\ri\,\chi_n(x,x';0)]\right|^2\nonumber\\
&\quad-\left\{\frac{[f(x)-f(x')]\,t}{\hbar}\right\}^2\,
\sum_n (E_n)^2\,a_n(x,x';0)\,\cos[\chi_n(x,x';0)]\nonumber\\
&\quad+\cdots
\end{align}
Both $\rho(x,x',n,n:0)$ and $A(x,x';0)$ are real and nonnegative for $x'=x$.
Hence $\chi_n$ vanishes for $x'=x$, so that up to second order in $(x'-x)$
one has
\begin{align}
|\gamma(x,x';t)|^2&=1+\frac{2\,[f(x)-f(x')]\,t}{\hbar}\,(x'-x)\,
\langle E\,\mathrm{D}\chi\rangle(x;0)\nonumber\\
&\quad-\left\{\frac{[f(x)-f(x')]\,t}{\hbar}\right\}^2\,(\Delta E)^2(x;0)
+\cdots,\\
&=1-(x'-x)^2\,\Bigg(2\,\frac{[(\rD f)(x)]\,t}{\hbar}\,\langle E\,\mathrm{D}\chi\rangle(x;0)
\nonumber\\
&\quad+\left\{\frac{[(\rD f)(x)]\,t}{\hbar}\,\right\}^2\,(\Delta E)^2(x;0)\Bigg)+\cdots
\end{align}
where $\rD$ stands for differentiation and
\begin{align}
\langle E\,\mathrm{D}\chi\rangle(x;0)&=\sum_n E_n\,a_n(x,x;0)\,\mathrm{D}_2\,\chi_n(x,x;0).\\
(\Delta E)(x;0)&=\left\{\sum_n (E_n)^2\,a_n(x,x;0)-\left[\sum_n E_n\,a_n(x,x;0)\right]^2\right\}^{1/2}.
\end{align}
$\rD_i$ is differentiation with respect to the $i$'th variable.
Of course $[a_n(x,x;0)]$'s are real nonnegative and add up to unity.

The above allows for a change in coherence which is first order in time, unless
$\chi_n$'s vanish. But if there is no first-order change, then the coherence
decreases initially. There are two time scales corresponding to the change in $|\gamma(x,x';t)|$:
\begin{align}
t_1&=\frac{h}{|(x'-x)^2\,[(\rD f)(x)]\,[\langle E\,\mathrm{D}\chi\rangle(x;0)]|}.\\
t_2&=\frac{h}{|(x'-x)\,[(\rD f)(x)]|\,[(\Delta E)(x;0)]}.
\end{align}
In terms of proper times, these become
\begin{align}
\tau_1&=\frac{h\,f(x)}{|(x'-x)^2\,[(\rD f)(x)]\,[\langle E\,\mathrm{D}\chi\rangle(x;0)]|}.\\
\tau_2&=\frac{h\,f(x)}{|(x'-x)\,[(\rD f)(x)]|\,[(\Delta E)(x;0)]}.
\end{align}
Both time scales are inversely proportional to $(\rD f)(x)$.

The origin of the first order change in coherence, could be traced back to (\ref{gd.19}).
It is seen that if $\chi_n$'s are initially zero, then $\phi_n$'s are initially zero
and the coherence is maximum (it decreases with time). If $\chi_n$'s are nonzero but
\textit{in the opposite phase} with the energy times the difference of the red shifts,
then $\phi_n$'s decrease with time, making the system more like the case of vanishing
$\chi_n$'s (vanishing $\phi_n$'s), hence increasing the coherence. It could also be noted
that $\phi_n$'s, even if initially are real don't remain so. So, if one begins with
a system with vanishing $\chi_n$'s (vanishing $\phi_n$'s), time evolution produces
(for small times) $\phi_n$'s which are \textit{in phase} with the energy times
the difference of the red shifts, hence decreasing the correlation further.
\section{Thermal equilibrium}
For a system localized at the point $x$, and having a local temperature $T(x)$,
the density matrix is
\begin{align}
\rho&=|x\rangle\langle x|\,\rho_\mathrm{in}[T(x)],\\
\intertext{where}
\rho_\mathrm{in}(T)&=\frac{1}{Z_\mathrm{in}(T)}\,
\sum_n |n\rangle\langle n|\,\exp\left(-\frac{E_n}{k_\mathrm{B}\,T}\right),\\
Z_\mathrm{in}(T)&=\sum_n\exp\left(-\frac{E_n}{k_\mathrm{B}\,T}\right).
\end{align}
If the system is delocalized, it is not the local Hamiltonian which is conserved.
Consider changes $\Delta _1$ and $\Delta_2$ in the local energies at the points
$x_1$ and $x_2$, respectively. One has
\begin{align}
\Delta_2&=-\frac{f(x_1)}{f(x_2)}\,\Delta_1.\\
\intertext{$\Delta S$, the change of entropy, would then be}
\Delta S&=\frac{\Delta_1}{T_1}+\frac{\Delta_2}{T_2}.\\
\intertext{A necessary condition for thermal equilibrium is that $\Delta S$
be zero. This results in}
[f(x_2)]\,T_2&=[f(x_1)]\,T_1.\\
\intertext{So a necessary condition for the thermal equilibrium would be
the existence of a global (position-independent) temperature $\bm{T}$ such that
the local temperatures satisfy}
T(x)&=\frac{\bm{T}}{f(x)}.
\end{align}
A density matrix corresponding to local temperatures $T(x)$ at each point $x$, is
not unique. It should satisfy
\begin{equation}
\rho(x,x;n,n')=A(x,x)\,\langle n|\rho_\mathrm{in}[T(x)]|n'\rangle.
\end{equation}
But this does not say anything about $\rho(x,x';n,n')$ with $x'\ne x$.
A simple choice is
\begin{align}\label{si}
\rho(x,x';n,n')&=\frac{A(x,x')}{Z_\mathrm{in}(\bm{T};x,x')}\,
\exp\left\{-\frac{E_n\,[f(x)+f(x')]}{2\,k_\mathrm{B}\,\bm{T}}\right\}\,\delta_{n\,n'},\\
\intertext{where}
Z_\mathrm{in}(\bm{T};x,x')&=\left\{Z_\mathrm{in}\left[\frac{\bm{T}}{f(x)}\right]\,
\,Z_\mathrm{in}\left[\frac{\bm{T}}{f(x')}\right]\right\}^{1/2}.
\end{align}
Such a choice results in
\begin{align}
\chi_n(x,x')&=0.\\
a_n(x,x')&=\frac{1}{Z_\mathrm{in}(\bm{T};x,x')}\,
\exp\left\{-\frac{E_n\,[f(x)+f(x')]}{2\,k_\mathrm{B}\,\bm{T}}\right\}.
\end{align}
The choice \Ref{si} for the density matrix at $t=0$, gives
\begin{align}
\gamma(x,x';t)&=\frac{Z_\mathrm{in}(\mathcal{T})}{Z_\mathrm{in}(\bm{T};x,x')},\\
\intertext{where}
\frac{1}{\mathcal{T}}&=\frac{f(x)+f(x')}{2\,\bm{T}}-\frac{k_\mathrm{B}\,[f(x)-f(x')]\,t}{\ri\,\hbar}.
\end{align}
One also notices that in this case,
\begin{align}
\langle E\,\mathrm{D}\chi\rangle(x;0)&=0,\\
(\Delta E)^2(x;0)&=k_\mathrm{B}\,\left[\frac{\bm{T}}{f(x)}\right]^2\,C(x),
\end{align}
where $C(x)$ is the local heat capacity at the point $x$, corresponding to
the internal degrees of freedom. For small times, the coherence deceases
quadratically in time, with a time scale
\begin{align}
\tau_2&=\frac{h\,[f(x)]^2}{|(x'-x)\,[(\rD f)(x)]|\sqrt{k_\mathrm{B}\,C(x)}\,\bm{T}},\\
&=\frac{h\,f(x)}{|(x'-x)\,[(\rD f)(x)]|\sqrt{k_\mathrm{B}\,C(x)}\,T(x)}.
\end{align}
For a system not in extreme conditions (not in strong gravitational fields),
one could approximate the above:
\begin{align}
\tau_2&\approx\frac{h}{k_\mathrm{B}\,T}\,\sqrt{\frac{k_\mathrm{B}}{C}}\,\frac{c^2}{g\,\Delta x},
\intertext{where $f$ is taken to be nearly one, so that the $x$-dependences are neglected except
for the derivative of $f$. $g$ is the gravitational acceleration, and use has been made
of the fact that $c^2\,(f-1)$ is the gravitational potential \cite{CW}, for example.
If the effective number of internal degrees of freedom (those which have been excited)
is neither small nor large, say considering a simple molecule in room temperature,
then $C$ is of the order of $k_\mathrm{B}$. Then at room tempertaure,}
\tau_2&\sim\frac{10^4\,\mathrm{m^2\,s^{-1}}}{g\,\Delta x}.
\intertext{which for a system of size around $1\,\mathrm{m}$ on earth results in}
\tau_2&\sim 10^3\,\mathrm{s}.
\end{align}
\section{Concluding remarks}
The effect of gravitational time dilation of the correlation between parts
of a small quantum system located at different points was studied. It was shown that
\begin{itemize}
\item External potentials do not affect the correlation.
\item There are in general two time scales for the change in correlation.
The first time scale is finite, only if there are non-real correlations between
different points. This time scale results in a change in the correlation
which is first order in time, and could lead to an increase or a decrease
in the correlation. The second time scale is always there, and results in
a change in the correlation which is second order in time, and always leads to
a decrease in the correlation.
\end{itemize}
\noindent
\textbf{Acknowledgement}: The work of AS and MK was supported by
the research council of the Alzahra University.

\end{document}